\documentclass{aa}
\usepackage{graphicx}
\usepackage{longtable}
\begin{document}
   \title{The relation between AGN hard X-ray emission and mid-infrared 
   continuum from ISO spectra: Scatter and unification aspects
   \thanks
   {Based on observations with ISO, an ESA project with instruments funded
   by ESA member states (especially the PI countries: France, Germany, the
   Netherlands, and the United Kingdom) with the participation of ISAS and
   NASA.}
   }

   \subtitle{}

   \author{D. Lutz \inst{1}
          \and
          R. Maiolino \inst{2}
          \and
          H.W.W. Spoon \inst{3}
          \and
          A.F.M. Moorwood\inst{4}
          }

   \offprints{D. Lutz}

   \institute{Max-Planck-Institut f\"ur extraterrestrische Physik, 
              Postfach 1312, 85741 Garching, Germany\\
              \email{lutz@mpe.mpg.de}
         \and
             Osservatorio Astrofisico di Arcetri, Largo E. Fermi 5, 50125
             Firenze, Italy\\
              \email{maiolino@arcetri.astro.it}
          \and
              Cornell University, Dept. of Astronomy, 219 Space Science 
              Building, Ithaca, NY 14853-6801, USA\\
              \email{spoon@isc.astro.cornell.edu}
        \and
             European Southern Observatory, Karl-Schwarzschild-Str. 2,
             85748 Garching, Germany\\
             \email{amoor@eso.org}
             }

   \date{Received 10/12/2003; accepted 26/01/2004}

   \abstract{We use mid-infrared spectral decomposition to separate the
   6$\mu$m mid-infrared AGN continuum from the host emission in the 
   ISO low resolution spectra of 71 active galaxies and
   compare the results to observed and intrinsic 2-10keV hard X-ray fluxes 
   from the
   literature. We find a correlation between mid-infrared luminosity and 
   absorption corrected hard X-ray luminosity, but the scatter is about 
   an order of magnitude, significantly larger than previously found
   with smaller statistics. Main contributors to this scatter are likely 
   variations in the geometry of absorbing dust, and AGN variability in 
   combination with non-simultaneous observations.
   There is no significant difference between Type 1 and 
   Type 2 objects in the average ratio of mid-infrared and hard X-ray emission,
   a result which is not consistent with the most simple version of a
   unified scheme in which an optically and geometrically thick torus
   dominates the mid-infrared AGN continuum.
   Most probably, significant  non-torus contributions to the AGN mid-IR 
   continuum are masking the expected difference
   between the two types of AGN. 
   }

\titlerunning{Mid-infrared and hard X-ray emission in AGN}
\authorrunning{D. Lutz et al.}

   \maketitle
%
%________________________________________________________________

\section{Introduction}
The mid-infrared and hard X-rays are two of the regions of the 
electromagnetic spectrum that are of particular interest for the study 
of active galactic nuclei (AGN). Hard X-rays, unless extremely obscured in 
fully Compton-thick objects, can provide a direct view to
the central engine, and are often believed to be a reasonable isotropic 
measure of the bolometric luminosity of the AGN. The nuclear
infrared continuum in Seyferts, in contrast, is due to AGN emission
reprocessed by dust, either in the putative torus or on somewhat larger
scales, e.g. inside the Narrow Line Region. The observed mid-infrared
emission is thus a function not only of the AGN luminosity but also 
of the distribution of the obscuring matter and of the viewing direction
of the observer. In the most simple form
this is due to the covering factor and distance of the obscuring dust,
but much more complex radiative transfer effects may occur in high
optical depth configurations (e.g. Pier et al. \cite{pier92}). 
In general terms, measurements
of the mid-infrared continuum in conjunction with hard X-ray observations 
can be thought of as testing unification scenarios for AGN. A tight
relation between the two quantities has recently been reported by
Krabbe et al. (\cite{krabbe01}) on the basis of mid-infrared imaging
of eight nearby Seyferts. Clavel et al. (\cite{clavel00}) have found a 
large difference between the equivalent widths of the mid-infrared aromatic
`PAH' emission features in Type 1 and 2 Seyferts. Under the assumption that 
the Type 1 and 2 subsamples are well matched in AGN luminosity and host
properties, they interpret this as an orientation dependent depression
of continuum in Seyfert 2s with respect to the host-related isotropic PAH
emission.

The issue is plagued, however, by the technical difficulty of isolating
the AGN mid-infrared continuum from the host galaxy emission. IRAS has 
readily detected large numbers of AGN, but the host contribution to 
these large beam infrared spectral energy distributions 
(e.g. Spinoglio et al. \cite{spinoglio95}) 
is not easy to quantify and significant in all but the very powerful AGN.
One way to address this difficulty is high spatial resolution imaging from
groundbased telescopes, e.g. in the L-  and M-bands (e.g. Alonso-Herrero
et al. \cite{alonso01}) or the N-band (e.g. Maiolino et al. \cite{maiolino95};
Krabbe et al. \cite{krabbe01}). This method produces reliable results in
cases of good surface brightness contrast between AGN and host, but can face 
ambiguities in cases where the AGN is surrounded by intense star formation 
on scales similar to the spatial resolution used, in particular if the
observations are diffraction limited by a moderate size telescope  
(e.g. \object{NGC\,6240}, \object{NGC\,4945}; Krabbe et al. \cite{krabbe01}).

We use the alternative approach of isolating the AGN mid-infrared
continuum {\em spectrally}, making use of the sizeable database of low 
resolution mid-infrared spectra of AGN that are a legacy of the Infrared Space
Observatory ISO. Low resolution spectra of galaxies can be decomposed into
three components (Laurent et al. \cite{laurent00}): A component dominated 
by the aromatic `PAH' features arising in photodissociation regions or the 
diffuse interstellar medium of the host, an
H\,II region very small grain continuum rising steeply towards wavelengths 
beyond 10$\mu$m,
and for active galaxies a typically flatter thermal AGN dust continuum. 
Starlight is 
unimportant except for quiescent objects like ellipticals or nearby spirals
with weak central star formation or AGN activity. 
The three components may also be obscured,
with the additional complication of ice features (Spoon et al. \cite{spoon02}).
A full spectral decomposition accounting for all these effects can be attempted
in cases of good signal-to-noise ratio and full wavelength coverage 
(e.g. Tran et al. \cite{tran01}, Spoon et al. \cite{spoon03}). 
Since most of our
data are for the restricted ISOPHOT range (5.8 to 11.8 $\mu$m) that limits the
accuracy of separating silicate absorption and PAH emission, and since
some spectra are of limited S/N, we follow a more straightforward approach.

In the range covered by the ISOPHOT spectra, the AGN emission is most
easily isolated shortwards of the complex of aromatic emission features
(Laurent et al. \cite{laurent00}). We determine a continuum at 6$\mu$m rest 
wavelength, and eliminate non-AGN emission that will in many cases be 
present in the fairly large beam. This is done by subtracting a star formation
template scaled with the strength of the aromatic `PAH' features arising in the
host or in circumnuclear star formation. This method does not require to
spatially resolve the AGN from the contaminating star formation, and will
face its limits only when trying to identify a weak AGN in the presence of
strong star formation or a stellar continuum that can be detectable for the
most nearby galaxies.
Our sample of 71 AGN is then used to quantify the relation of mid-infrared and
X-ray emission at significantly better statistics than previously possible. 

\section{Data}

We have included in our sample those Seyferts or Quasars with low
resolution spectra in the ISO archive for which hard X-ray observations
were available in the literature. Since our main goal is a comparison of 
thermal infrared
emission with hard X-rays we have excluded objects believed to have a 
significant synchrotron contribution in the mid-infrared (e.g. Cen A, 3C273). 
Of the Ultraluminous Infrared Galaxies for which a 
significant number of ISO spectra is available, we have restricted ourselves to
a number of the brightest objects with clear and undisputed AGN contribution. 

\subsection{Infrared Spectroscopy}
Most of the spectra are from a rereduction of chopped ISOPHOT-S spectra of
galactic nuclei in the ISO archive (Spoon et al. 2004, in preparation).
The reduction was done
using PIA\footnote{PIA is a joint development by the ESA Astrophysics division
and the ISOPHOT consortium} version 9.0.1. 
Steps in the data reduction included: 1)
deglitching on ramp level. 2) subdivision of ramps in four sections
of 32 non destructive read-outs. 3) ramp fitting to derive
signals. 4) masking of bad signals by eye-inspection. 5) kappa sigma
and min/max clipping on remaining signal distribution. 6)
determination of average signal per chopper plateau. 7) masking or
correction of bad plateaux by eye-inspection. 8) background
subtraction by subtracting off-source from on-source plateaux. 9) finally, flux
calibration, using the signal dependent spectral response function.
The absolute calibration is accurate to within 20\%. The ISOPHOT-S aperture
is $24\arcsec\times 24\arcsec$ and in many cases includes noticeable host 
emission. A few spectra are based on different reductions. Circinus and
NGC\,1068 were observed in staring mode, and for NGC\,4945 the ramp division of
step 2) used only two sections. Mrk\,1014 was reduced with the earlier PIA 
7.2 and scaled to the PIA 9.0.1 flux scale. For NGC\,1808 we have used an 
ISOCAM-CVF spectrum (Laurent et al. \cite{laurent00} and priv. communication).

\begin{figure}
\includegraphics[width=\columnwidth]{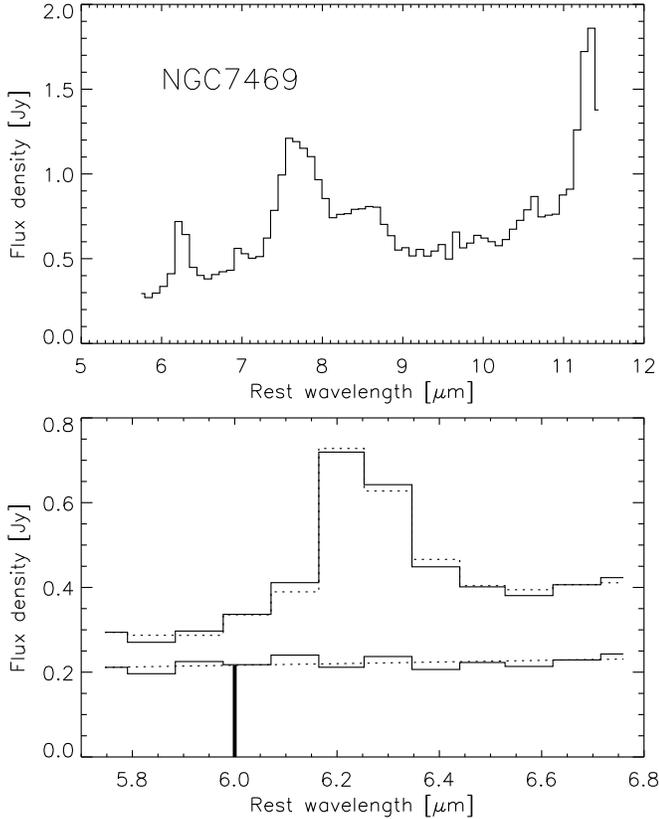}
\caption{Illustration of the decomposition used to isolate the AGN
continuum. Top panel: ISOPHOT-S spectrum of the mixed AGN/starburst galaxy 
NGC\,7469, showing strong PAH emission as well as an elevated continuum.
Bottom panel: Cutout of the region around the 6.2$\mu$m PAH feature.
Top continuous line = observed spectrum. Top dotted line = fit by the
sum of the rebinned SWS M82 PAH spectrum and a linear AGN continuum.
Bottom dotted line = fitted AGN continuum. Bottom continuous line =
difference of observed spectrum and fitted PAH component. The thick
vertical line indicates the AGN 6$\mu$m continuum, obtained
by evaluating the fitted continuum at that wavelength. An additional 
correction is applied to this value to account for differences between
M\,82 and an average starburst.}
\label{fig:decomposedemo}
\end{figure}

We derive the 6$\mu$m AGN continuum by a decomposition over a range
from 5.5$\mu$m rest wavelength (or the minimum rest wavelength covered) 
to 6.85$\mu$m. With the exception of the 6.2$\mu$m PAH feature
(also associated with a continuum-like `plateau'), this region is 
relatively free of spectral structures
or strong emission lines. Absorption features from ice (6$\mu$m) and 
hydrocarbons (6.8$\mu$m) are found in highly obscured objects but are mostly
undetected 
in our AGN sample (see Spoon et al. \cite{spoon02}). We can thus fit
the spectrum by the superposition of a star formation component dominated
by the 6.2$\mu$m PAH feature and a simple linear approximation for the AGN
continuum. For the star formation component, we use the SWS spectrum of
M\,82 (Sturm et al. \cite{sturm00}), rebinned to the exact sampling
of the individual low resolution spectra. Fig.~\ref{fig:decomposedemo}
illustrates this using the example of NGC\,7469 which is well known to host
both a powerful Type 1 AGN and a strong circumnuclear star forming ring
(e.g., Mazzarella et al. \cite{mazzarella94}, Genzel et al. \cite{genzel95}).
The correction for `continuum' associated with the aromatic emission is
relatively modest in this object, but more than half in some Seyferts in 
our sample. The results of the fit procedure are a 6$\mu$m continuum
as illustrated in Fig.~\ref{fig:decomposedemo} and an average flux density
of the PAH component over the rest wavelength range 6.1 to 6.35$\mu$m, both
including a 1+z correction so that the rest frame spectrum $\nu f_\nu$ 
matches the observed frame one.

The SWS spectrum of M\,82 is technically well suited for this decomposition
but represents just a single object that may be not fully representative
for star forming objects in general, given a certain dispersion in relative
importance of `PDR' and `HII' components in the sense of the model of
Laurent et al. (\cite{laurent00}). We have thus applied the decomposition
also to a number of bona fide star forming galaxies (NGC\,23, NGC\,232, 
NGC\,520, IC\,342, NGC\,3256, M\,83, NGC\,5653, NGC\,6090, NGC\,6946, 
NGC\,7252, NGC\,7552), excluding starbursts with clear evidence for 
an additional minor AGN (e.g. NGC\,3690, Della Ceca et al. 
\cite{dellaceca02}). Indeed, M\,82 is found to have relatively low
6$\mu$m continuum compared to most of these objects. Averaging over all 
these star forming objects including 
M82 we find a mean residual 6$\mu$m continuum which is 0.096 times the average
6.1 to 6.35$\mu$m PAH flux density, with a dispersion of 0.085. 
We consider both this average value and the dispersion in our analysis of
AGN. The 6$\mu$m AGN continua listed in Table~\ref{tab:fluxes} 
are the values obtained after subtracting from the direct fit result 
0.096 times the average 6.1 to 6.35$\mu$m PAH flux density.
Error estimates for the 6$\mu$m continuum are the quadratic sum of two
components. The first is a measurement error based on individual pixel noise 
derived from the dispersion in the difference of observation and fit. The
second is 0.085 times the average 6.1 to 6.35$\mu$m PAH flux density, thus 
considering the dispersion in the properties of the comparison star forming 
galaxies. Either component is found to dominate for part of the AGN sample,
depending on S/N and the relative importance of the PAHs.
Limits quoted in Table~\ref{tab:fluxes} are 3$\sigma$ limits based on these
considerations.

We ascribe the 6$\mu$m continuum derived in this way to an AGN. This is not 
strictly correct in all cases, as very intense star forming environments 
outside of the parameter range covered by our comparison objects also can 
produce a stronger continuum with relatively
weak PAH at these wavelengths. This is true, for example, for the
obscured region in SBS 0335-052 (Thuan et al. \cite{thuan99}), the
mid-infrared peak in the Antennae (Mirabel et al. \cite{mirabel98}),
and likely the circumnuclear region of Arp 220 (Spoon et al. \cite{spoon03}).
We consider this of minor importance, however, 
for our sample which does not include star forming dwarfs, and only some
of the best established AGN among the ULIRGs. 

In some of the weaker AGN in nearby galaxies, photospheric emission from the 
central old stellar population
may contribute significantly to the measured 6$\mu$m continuum. The 6$\mu$m 
stellar continuum can be extrapolated from the stellar K-band continuum.
From four ellipticals without PAH emission (NGC\,3379, NGC\,4374, NGC\,2300,
NGC\,4649; Lu et al. \cite{lu03}, Xilouris et al. \cite{xilouris03}, S. Madden
priv. comm.) we estimate a scaling $S_{6\mu m}\sim 0.19\times S_{2.2\mu m}$. 
We cannot 
directly apply this extrapolation to our sample objects, since it applies 
only to the stellar K-band, and decompositions of the K-band continuum inside
our aperture into stellar and AGN are usually not available. We have verified,
however, that the conclusions reached in Sect. 3 are robust to stellar 
continuum contributions. For this purpose, we have repeated as an extreme
assumption our 
analysis after subtracting directly from the 6$\mu$m continuum obtained in the 
spectral decomposition 0.19 times the total K-band 
continuum in an ISOPHOT-S aperture, which can be extrapolated with modest
uncertainty from 2MASS data accessible in NED. The results given in
Table 2 did not change significantly. We did not adopt
these values, however, since they imply a systematic overcorrection for the
many AGN dominated objects. Instead, we mark in Table~\ref{tab:fluxes} 
the few cases where a strong stellar contribution to our measured 6$\mu$m continuum is likely.

We do not attempt a correction of the mid-infrared fluxes for foreground 
extinction. Published extinction values for those objects refer
to different tracers, and their applicability to the region dominating the
mid-infrared flux is uncertain. Since the extinction at 6$\mu$m is 1/20 or less
of the visual extinction, typical A$_V$ values from optical studies are
irrelevant in any case. In the discussion of individual objects in the
Appendix, we also mention NGC\,6240 and NGC\,4945  where obscuration of the
mid-infrared AGN continuum is likely significant.

\subsection{X-ray emission}

\begin{figure}
\includegraphics[width=\columnwidth]{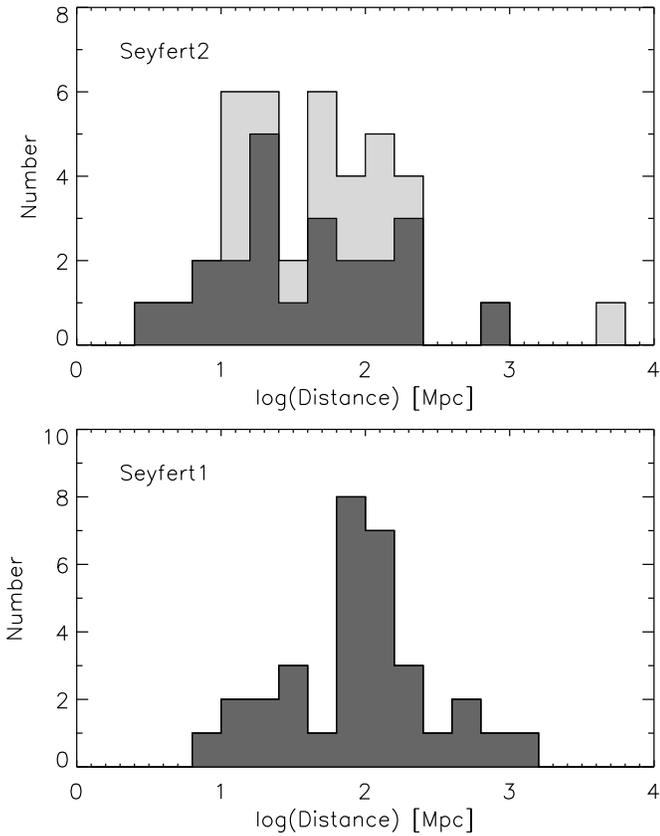}
\caption{Distance distribution for the objects in our sample, separated
by Type 1 (1 to 1.5) and 2 (1.8 to 2) Seyferts. Light shading identifies those
Seyfert 2's where no absorption correction to the AGN hard X-ray emission
was possible. We adopted distances derived assuming H$_0$=75, q$_0$=0.5 
except for M51 (8.4Mpc), M81 (3.63Mpc) and Circinus (4Mpc).}
\label{fig:disthisto}
\end{figure}

We have compiled the X-ray fluxes in the 2-10keV range from various 
literature sources using data from different satellites (e.g. ASCA, BeppoSAX,
Chandra, XMM). For Seyfert 2s for which the absorbing column density 
$N_H$ could be measured (most cases), the authors generally provide
reliable measurements of the instrinsic, absorption-corrected hard X-ray 
flux. The intrinsic X-ray flux cannot be recovered when only a lower limit to 
$N_H$ is inferred; this is the case for totally Compton thick nuclei with
$N_H>10^{25}$cm$^{-2}$, which are absorbed at all energies, or for mildly 
Compton thick nuclei with $N_H>10^{24}$cm$^{-2}$ which are lacking data at 
E$>$10keV, i.e. for which any transmitted flux at high energy cannot be probed.
For Seyfert 1 galaxies without published X-ray column we have assumed that
observed and intrinsic flux can be safely adopted to be equal because
of typically low absorbing columns. Figure~\ref{fig:disthisto} shows 
the distance distributions
both for the full sample and for the subsample with absorption corrected
X-ray fluxes.

\begin{table*}
\caption{Infrared and hard X-ray fluxes for our sample. The 6$\mu$m flux
refers to the AGN continuum exclusively and is derived from ISO 
low resolution spectra as described in the text. Hard X-ray data
have been taken from the references listed.}
\begin{tabular}{lrlrrrrrl}\hline
Name     &cz$_h$&Type &S$_{6\mu m}$&S$_{PAH}$&
F$_{2-10keV,obs}$&$N_H$&F$_{2-10keV,int}$&
Reference\\
         &km/s  &     &Jy          &Jy       &10$^{-11}$erg\,s$^{-1}$cm$^{-2}$&
10$^{20}$cm$^{-2}$&10$^{-11}$erg\,s$^{-1}$cm$^{-2}$\\ \hline
Mrk\,334 &  6582&Sy1.8&   0.0357&   0.0054&   0.80&   ? &?$^5$&Polletta96\\ 
Mrk\,335 &  7730&Sy1  &   0.1508&$<$0.0302&   0.96&  4.0&0.96&Gondoin02\\
NGC\,34  &  5931&Sy2  &$<$0.0804&   0.1810&$<$0.39&   ? & ?  &Polletta96\\
I\,Zw\,1 & 18342&Sy1  &   0.1859&$<$0.0325&   0.68&  3.0&0.68&Reeves00\\
Fairall\,9&14095&Sy1  &   0.1386&$<$0.0188&   1.53&  2.3&1.53&Gondoin01a\\
NGC\,526A&  5725&Sy1.9&   0.1114&$<$0.0182&   1.80&200.0&1.93&Landi01\\
Mrk\,359 &  5215&Sy1  &   0.0432&   0.0318&   0.74&  4.8&0.74&OBrien01\\
Mrk\,1014& 48900&Sy1  &   0.0218&$<$0.0222&   1.24&   ? &1.24$^4$&Boller02\\
Mrk\,590 &  7910&Sy1  &   0.1132&   0.0232&   1.97&   ? &1.97$^4$&Malizia99\\
NGC\,1068&  1148&Sy2  &  11.2195&   1.2491&   0.35&$>$1e5&?$^5$&Matt97\\
4U0241+61& 13200&Sy1  &   0.2673&$<$0.0269&   3.57&150.0&3.82&Malizia97\\
NGC\,1097&  1275&Sy1  &$<$0.3087&   0.8532&   0.17&  6.0&0.17&Iyomoto96\\
NGC\,1144&  8648&Sy2  &$<$0.0135&$<$0.0119&$<$11.00&  ? &  ? &Polletta96\\
NGC\,1365&  1636&Sy1.8&$<$0.5582&   1.2006&   0.66&4000.0&2.31&Risaliti00\\
NGC\,1386&   868&Sy2  &   0.1684&$<$0.0349&   0.02&   ? &?$^5$&Maiolino98\\
Mrk\,618 & 10658&Sy1 &$^3$0.0179&$<$0.0287&   0.70&  3.0&0.70&Rao92\\
04385-0828& 4527&Sy2  &   0.2205&$<$0.0119&   1.80&   ? &?$^5$&Polletta96\\
NGC\,1667&  4547&Sy2$^1$&$<$0.0542&   0.1360&  0.003&$>$1e4?&?$^5$&Bassani99\\
NGC\,1808&  1000&Sy2$^1$&$<$0.5108&   1.9983&   0.09&320.0&0.11&Bassani99\\
Ark\,120 &  9682&Sy1  &   0.1106&   0.0215&   3.58&   ? &3.58$^4$&Nandra94\\
05189-2524&12760&Sy2  &   0.2114&   0.0416&   0.36&800.0&0.95&Severgnini01\\
MCG8-11-11& 6141&Sy1.5&   0.1890&$<$0.0207&   5.66&$<$2.0&5.66&Perola00\\
Mrk\,3   &  4050&Sy2  &   0.0931&$<$0.0177&   0.65&1.1e4 &4.08&Cappi99\\
Mrk\,6   &  5536&Sy1.5&   0.1424&$<$0.0183&   1.20&1000.0&2.00&Immler03\\
PG0804+761&30000&Sy1  &   0.0778&$<$0.0103&   0.88&$<$40.0&0.88&George00\\
M\,81    &   -34&Sy1.8&$^3$0.3835&$<$0.0278&   1.50&   9.4&1.50&Bassani99\\
NGC\,3227&  1157&Sy1.5&   0.1518&   0.0862&   0.75& 650.0&0.76&Gondoin03a\\
UGC\,6100&  8778&Sy2  &$<$0.0186&   0.0249&$<$1.14&    ? &  ? &Polletta96\\
NGC\,3516&  2649&Sy1.5&   0.2111&$<$0.0635&   4.41&   4.0&4.41&Guainazzi01\\
NGC\,3783&  2550&Sy1  &   0.2346&$<$0.0210&   8.50& 300.0&8.50&Blustin02\\
NGC\,3982&  1109&Sy2  &$<$0.0592&   0.1316&$<$0.42&    ? &  ? &Polletta96\\
NGC\,4051&   725&Sy1.5&   0.2014&$<$0.0768&   1.60&$<$1.0&1.60&Reynolds97\\
NGC\,4151&   995&Sy1.5&   0.9175&$<$0.0455&   9.00& 300.0&9.00&Yang01\\
Mrk\,766 &  3876&Sy1.5&   0.1415&   0.0354&   3.00&  80.0&3.00&Matt00\\
NGC\,4388&  2524&Sy2  &   0.1298&   0.1213&   1.20&4300.0&4.30&Bassani99\\
NGC\,4507&  3538&Sy2  &   0.1954&   0.0625&   2.10&2900.0&7.03&Bassani99\\
NGC\,4579&  1519&Sy1.9&$^3$0.0966&$<$0.0356&  0.44& 240.0&0.44&Bassani99\\
NGC\,4593&  2698&Sy1  &   0.1836&$<$0.0271&   3.71&   2.0&3.79&Perola02\\
NGC\,4945&   560&Sy2$^2$&$<$0.9092& 2.3230&   0.54&4.0e4 &19.00&Guainazzi00\\
NGC\,5033&   875&Sy1.9&$^3$0.0691&   0.1588&   0.55& 290.0&0.55&Bassani99\\
M\,51    &   463&Sy2  &$^3$0.0880&  0.1267&   0.03&5.6e4 &1.28&Fukazawa01\\
NGC\,5273&  1089&Sy1.9&   0.0216&$<$0.0190&$<$1.17&    ? &  ? &Polletta96\\
Mrk\,273 & 11132&Sy2  &   0.0725&   0.0617&   0.06&4900.0&0.35&Bassani99\\
IC4329A  &  4813&Sy1  &   0.4723&$<$0.0250&  16.40&  42.0&16.80&Gondoin01b\\
Mrk\,463 & 14904&Sy2  &   0.2518&$<$0.0304&   0.04&1600.0&0.09&Bassani99\\
Circinus &   439&Sy2  &   4.6688&   4.7199&   1.40&3.6e4 &40.00&Matt99\\
NGC\,5506&  1853&Sy1.9&   0.6607&   0.0665&   8.38& 340.0&10.80&Bassani99\\
IC\,4397 &  4419&Sy2  &$<$0.0228&   0.0587&   2.23&    ? &?$^5$&Polletta96\\
NGC\,5548&  5149&Sy1.5&   0.1302&   0.0267&   4.30&  51.0&4.30&Reynolds97\\
NGC\,5643&  1199&Sy2  &   0.0610&   0.0926&   0.13&$>$1e5&?$^5$&Maiolino98\\
NGC\,5674&  7474&Sy1.9&$<$0.0255&   0.0562&   0.84& 200.0&1.28&Bassani99\\
Mrk\,841 & 10852&Sy1.5&   0.0595&   0.0190&   1.00&    ? &1.00$^4$&Reynolds97\\
\hline
\end{tabular}
\label{tab:fluxes}

$^1$ Uncertain classification, see appendix.\\
$^2$ `Elusive' AGN without NLR detection but listed as type 2 according 
to the X-ray properties.\\
$^3$ Stellar continuum may contribute a large part of the measured 6$\mu$m
continuum.\\
$^4$ Assuming negligible absorption for Type 1 objects without $N_H$ 
measurement.\\
$^5$ Intrinsic flux cannot be recovered for Type 2 objects without $N_H$
measurement or with lower limits to $N_H$.
\end{table*}

\begin{table*}
\addtocounter{table}{-1}
\caption{continued}
\begin{tabular}{lrlrrrrrl}\hline
Name     &cz$_h$&Type &S$_{6\mu m}$&S$_{PAH}$&
F$_{2-10keV,obs}$&$N_H$&F$_{2-10keV,int}$&
Reference\\
         &km/s  &     &Jy          &Jy       &10$^{-11}$erg\,s$^{-1}$cm$^{-2}$&
10$^{20}$cm$^{-2}$&10$^{-11}$erg\,s$^{-1}$cm$^{-2}$\\ \hline
NGC\,5929&  2492&Sy2 &$^3$0.0171&$<$0.0273&$<$0.79&    ? &  ? &Polletta96\\
15307+3252&277700&Sy2 &   0.0073&$<$0.0023&$<$0.007&   ? &  ? &Ogasaka97\\
PG1613+658&38700&Sy1  &   0.0394&$<$0.0206&   0.69&   3.0&0.69&Lawson97\\
NGC\,6240&  7339&Sy2  &   0.0698&   0.2093&   0.19&2.2e4 &14.0&Vignati99\\
Mrk\,507 & 16758&Sy2  &   0.0094&   0.0141&   0.05&  34.0&0.05&Bassani99\\
H1821+643& 89038&Sy1  &   0.0972&$<$0.0114&   2.34&$<$1.8&2.34&Reeves00\\
ESO141-G55&10793&Sy1  &   0.0862&$<$0.0238&   2.65&   5.5&2.65&Gondoin03b\\
19254-7245&18500&Sy2  &   0.0786&$<$0.0593&   0.02&$>$1e4&?$^5$&Braito03\\
Mrk\,509 & 10312&Sy1  &   0.1226&   0.0345&   5.66&$<$1.0&5.66&Perola00\\
PKS2048-57& 3402&Sy2  &   0.2581&   0.0153&   1.20&2400.0&3.00&Bassani99\\
PG2130+099&18630&Sy1  &   0.0860&$<$0.0227&   0.53&$<$19.0&0.53&Lawson97\\
NGC\,7213&  1792&Sy1.5&$^3$0.1470&  0.0181&   3.66&    ? &3.66$^4$&Malizia99\\
NGC\,7314&  1422&Sy1.9&   0.0442&$<$0.0384&   3.56& 120.0&4.05&Bassani99\\
Ark\,564 &  7400&Sy1  &   0.0652&   0.0134&   2.00&  10.0&2.00&Turner01\\
NGC\,7469&  4892&Sy1  &   0.1788&   0.3873&   2.90&$<$0.4&2.90&Reynolds97\\
23060+0505&52200&Sy2  &   0.0963&$<$0.0301&   0.15& 900.0&0.25&Brandt97\\
NGC\,7582&  1575&Sy2  &   0.1827&   0.6378&   1.55&1200.0&2.72&Bassani99\\
NGC\,7674&  8713&Sy2  &   0.1541&   0.0594&   0.05&$>$1e5&?$^5$&Bassani99\\
NGC\,7679&  5138&Sy1  &$<$0.0827&   0.3209&   0.60&   5.0&0.60&DellaCeca01\\   
\hline
\end{tabular}
\label{tab:fluxescontinued}
\end{table*}

\section{Results}

\begin{figure}
\includegraphics[width=\columnwidth]{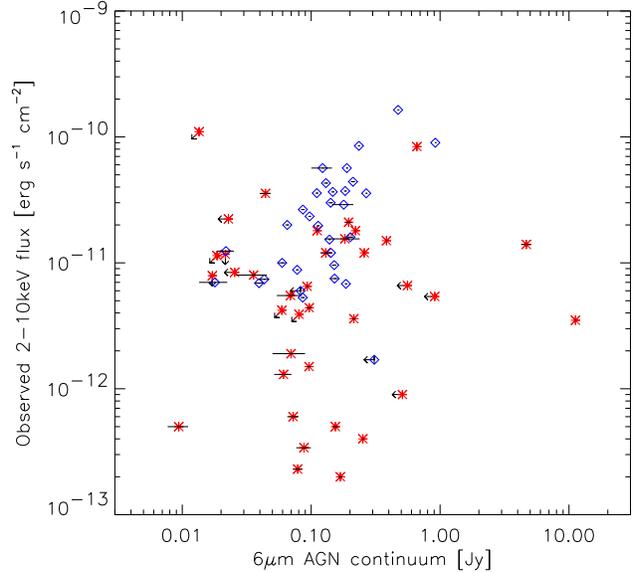}
\caption{Observed hard X-ray fluxes vs. 6$\mu$m AGN continuum for the sample 
galaxies. Diamonds indicate Type 1 Seyferts (including objects classified
as types 1, 1.2, 1.5), asterisks indicate Type 2 Seyferts (including objects 
classified as types 1.8, 1.9, 2). Thin horizontal lines, in most cases smaller
than the symbols, indicate the 1$\sigma$ uncertainty of the 6$\mu$m 
detections.}
\label{fig:obsxvsirflux}
\end{figure}

\begin{figure}
\includegraphics[width=\columnwidth]{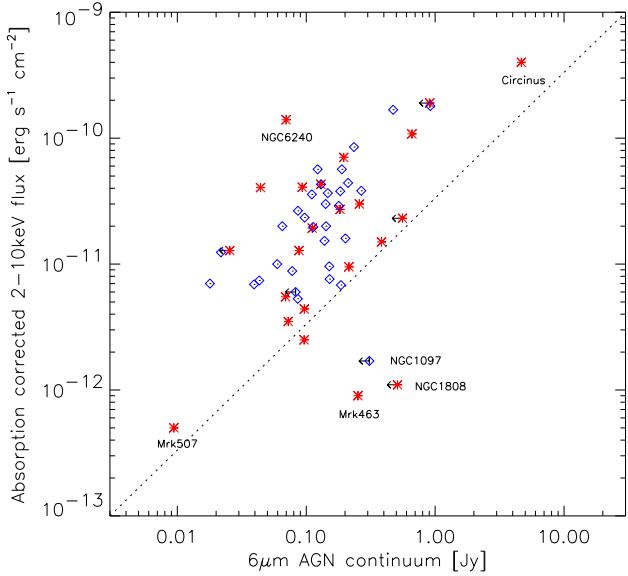}
\caption{Hard X-ray fluxes corrected for absorption vs. 6$\mu$m AGN 
continuum for those objects where an absorption corrected X-ray flux was
available. Symbols are as in Fig.~\ref{fig:obsxvsirflux}. The dotted line
indicates slope 1, it is not a fit.}
\label{fig:corrxvsirflux}
\end{figure}

\begin{figure}
\includegraphics[width=\columnwidth]{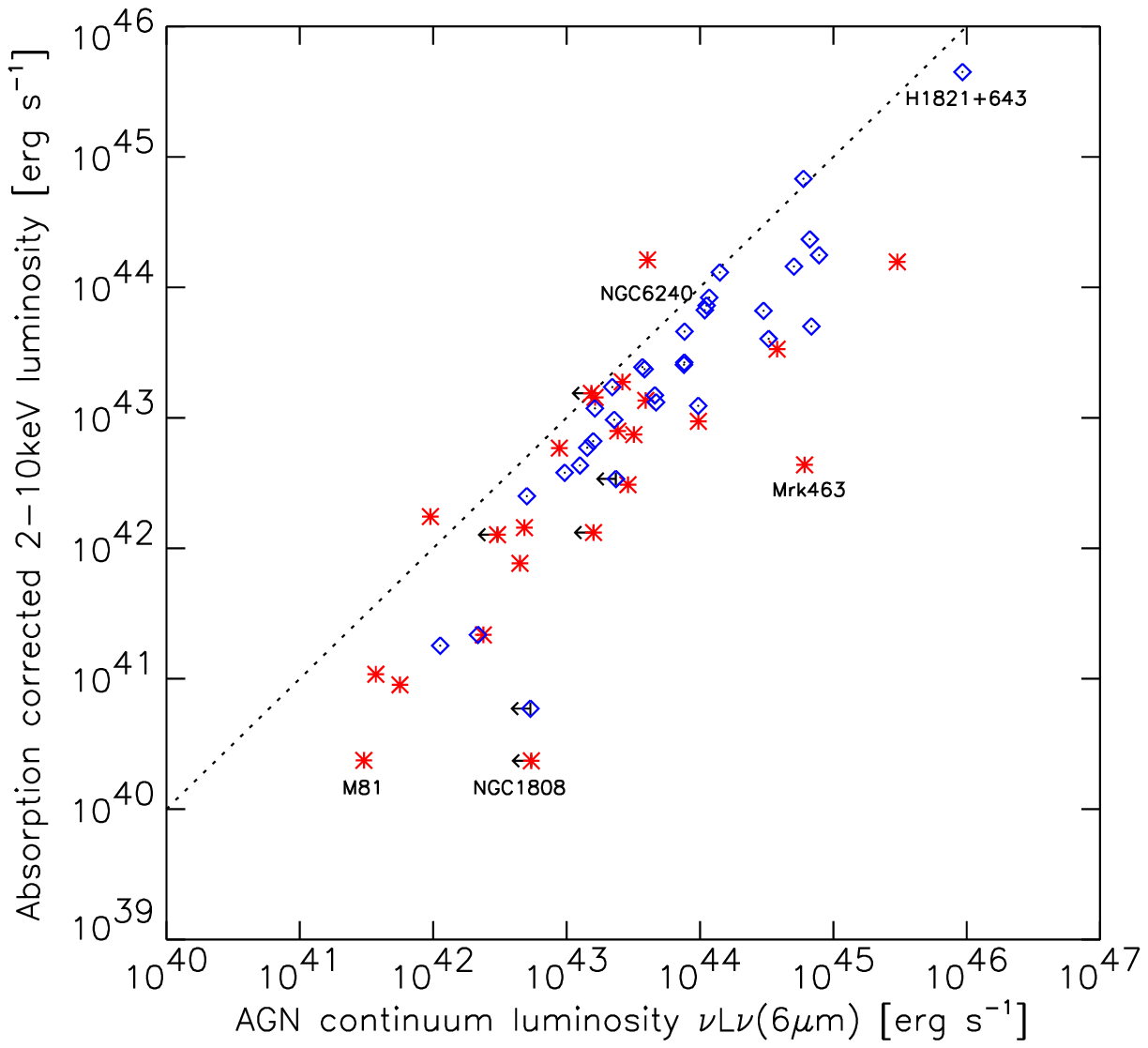}
\caption{Hard X-ray luminosities corrected for absorption vs. 6$\mu$m AGN 
continuum luminosities for those objects where an absorption corrected 
X-ray flux was
available. Symbols are as in Fig.~\ref{fig:obsxvsirflux}. The dotted line
indicates slope 1, it is not a fit.}
\label{fig:corrxvsirlum}
\end{figure}

The observed mid-infrared and hard X-ray fluxes are listed in 
Table~\ref{tab:fluxes} and summarized in 
Fig.~\ref{fig:obsxvsirflux}. The diagram shows the expected separation
between Type 1 and Type 2 Seyferts, hard X-ray emission in the latter often
being strongly absorbed. More interesting is the comparison with absorption
corrected X-ray fluxes shown in Fig.~\ref{fig:corrxvsirflux}. 
A correlation between hard X-rays and infrared continuum is indicated
in this diagram but with considerable scatter that is not due to
observational error in the mid-IR spectra. There is no offset between 
Type 1 and 2 objects any more.
Fig.~\ref{fig:corrxvsirlum} compares mid-infrared
and X-ray luminosities. Hard X-ray and mid-infrared luminosity correlate
over four orders of magnitude in luminosity, the ratio of the two
shows no clear trend with luminosity. The data are consistent
with the same mean ratio, with considerable dispersion, over the entire
range of luminosity.

\subsection{The scatter in the relation between mid-infrared continuum and 
hard X-ray emission}

\begin{figure}
\includegraphics[width=\columnwidth]{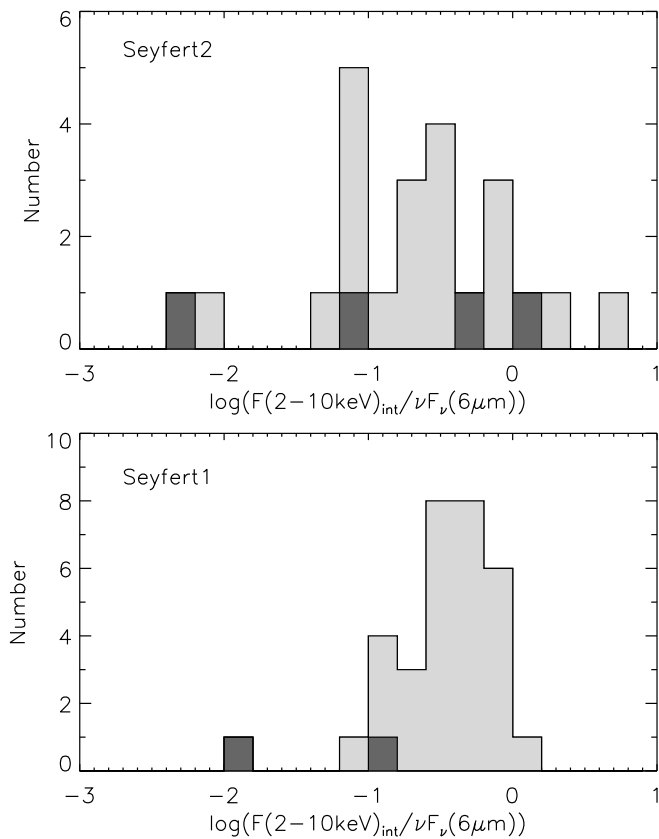}
\caption{Histograms showing for the two Seyfert types the logarithm of the
ratio between intrinsic hard X-ray flux and 6$\mu$m AGN continuum. Dark 
shadings indicates objects with upper limits for the 6$\mu$m continuum, i.e. 
lower
limits for the ratio shown.}
\label{fig:ratiohistogram}
\end{figure}

Fig.~\ref{fig:ratiohistogram} shows histograms for the ratio of 
hard X-ray flux and mid-infrared AGN continuum, separated by Seyfert type.
The dispersion is considerable, the ratio for Seyfert 1s varying over an 
order of magnitude. Ratios for Type 2 Seyferts scatter yet wider. This may be
 partly due to inaccuracies in the absorption correction
in the X-ray data. The current data do not allow to determine with
certainty whether there is in reality a larger spread for Type 2 than for
Type 1. The two outlying objects Mrk\,463 and NGC\,6240 are briefly
discussed in the Appendix. The spread observed in our 
sample is significantly larger than in the imaging study of Krabbe et al.
(\cite{krabbe01}), where with the exception of the outlier NGC\,6240
the ratio of intrinsic hard-X and mid-IR emission varies by just a 
factor of 3-4.
This was probably a fortuitious effect of the small sample including $\sim$10
times less objects than the present study. Given the many factors of AGN 
spectral energy distribution and geometry relating the hard X-ray flux and 
the dust reradiation of AGN emission in the mid-infrared, the larger 
scatter is not surprising. Another significant contribution to the spread
is likely due to AGN variability. The X-ray and infrared measurements 
summarized in Table~\ref{tab:fluxes} were usually
taken several years apart. Even in case of simultaneous observations,
short term variability of the central engine may contribute to the scatter 
because of different time response and time averaging effects for the X-rays 
and for the larger scale dust.  
 
\subsection{No significant difference between Seyfert 1 and Seyfert 2 types}

A surprising result given the reasonable statistics of the present sample
is the failure to detect the  difference between Type 1 and Type 2 Seyferts
that is predicted by the most straightforward versions of unified models.
The median log(F(2-10keV)/$\nu$F$_\nu$(6$\mu$m)) is -0.41 for Type 1
objects and -0.63 for Type 2. This would not change strongly replacing
the few lower limits by detections (Fig.~\ref{fig:ratiohistogram}). The 
key point is
not the small difference found, which might even be reduced slighlty
if some of the Sy2 limits were replaced by detections or with full correction 
for stellar continuum, it is the failure to detect
a strong and significant difference in the opposite direction. If 
the mid-infrared
continuum in Seyfert 2s is suppressed by a factor of $\sim$8 
(Clavel et al. \cite{clavel00}), the hard X/IR ratio should be higher
by the same factor compared to Seyfert 1s. While our targets are from a 
heterogeneous set of ISO observing
programs, they do not represent a preselection by mid-IR flux which may
affect such a comparison. The main observing programs involved
did not invoke such a selection. A major part, e.g., is formed by objects from
the CfA sample (Huchra \& Burg \cite{huchra92}). Also, for example, most
AGN from the hard X-ray selected sample  of Piccinotti et al. 
(\cite{piccinotti82}) are included. As discussed by Maiolino \& Rieke
(\cite{maiolino95b}), samples like the CfA one may be biased against 
obscured objects and thus not reproduce the real fractions of Seyfert types.
Our analysis normalizing to intrinsic X-rays should be robust to such effects
as long as reaching lower but still significant numbers of obscured systems. 

In unified schemes (e.g. Antonucci et al. \cite{antonucci93}), the difference
between Seyfert types is due to effects of viewing intrinsically similar
objects from different directions, because an anisotropic distribution of
absorbing material (e.g. the `torus') absorbs, scatters, and reprocesses
the direct AGN light. In the most simple version, a central very small
source (also emitting the X-rays) illuminates a torus-like geometrically
and optically thick dusty structure. Radiative transfer models of the
emission from a geometrically and optically thick torus (e.g. Pier et al. 
\cite{pier92}, Efstathiou \& Rowan-Robinson \cite{efstathiou95}, Granato et al.
\cite{granato97}, Nenkova et al. \cite{nenkova02}) predict a strong anisotropy 
of the mid-infrared emission. The unified model in this case 
clearly predicts for Seyfert 2 galaxies a higher ratio of absorption 
corrected X-ray emission and mid-infrared continuum.

In a simple unified scheme invoking a torus, the ratio of intrinsic hard X-ray
to observed mid-IR
emission will increase from a fully face-on to a fully edge-on
view. The difference between Type 1s and Type 2s could hence be masked if the
Type 2s in the sample were preferentially objects with still relatively low 
obscuring columns, likely
seen at intermediate angles. While we had to exclude fully Compton-thick 
Sy2s from the comparison because of the impossibility to derive an 
absorption corrected X-ray flux, they form well below half of the Sy2 
sample (5 of 38, more could be among the 11 Type 2 objects where the absorbing 
column is undetermined or only a limit to the 2-10keV flux available). 
Abandoning for a moment the normalization to X-rays, we have compared
the average ratio of 6$\mu$m AGN continuum to 6.2$\mu$m PAH feature  
for the Seyfert 2's without intrinsic X-ray flux in our sample
and the ones where an intrinsic flux could be derived. Both using the
measurements/limits in Table~\ref{tab:fluxes} and using the raw measurements,
the median AGN continuum to PAH ratio is only insignificantly ($\sim$20\%) 
smaller for the 16 Type 2 objects without intrinsic X-ray flux.
With all caveats of normalizing to the PAHs rather than to intrinsic X-rays, 
this tentatively indicates that
in a complete Sy2 sample including the fully Compton-thick objects
the mid-infrared AGN continuum will not be much lower.  

\begin{figure}
\includegraphics[width=\columnwidth]{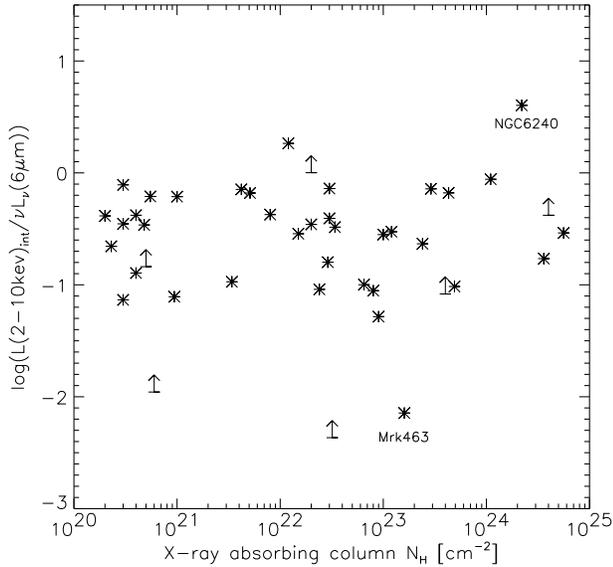}
\caption{Ratio of intrinsic hard X-ray flux and 6$\mu$m AGN continuum, plotted
as a function of X-ray absorbing column.}
\label{fig:ratiovscolumn}
\end{figure}

More insight can be obtained from plotting the ratio of 
X-ray and mid-infrared emission versus the X-ray absorbing column 
(Fig.~\ref{fig:ratiovscolumn}). Column densities above 
10$^{23}$cm$^{-2}$ are well populated but do not show the upturn in the
hard X/Mid-IR ratio expected at high columns in the simple unified scheme.
In a naive foreground screen the optical depth at 6$\mu$m exceeds one
already below 10$^{23}$cm$^{-2}$. Radiative transfer in different
more realistic `torus' geometries (see the large range
in the models cited above) will predict different magnitudes of this
effect and place it at different column densities. 
For example, a strong and fairly sharp upturn at 
$N_H\sim 6\times 10^{23}$cm$^{-2}$ is expected for the best fitting 
tapered disk model 5e of Efstathiou \& Rowan-Robinson (\cite{efstathiou95}).
Placing predictions of future torus models on a fully empirical diagram like 
Fig.~\ref{fig:ratiovscolumn} should provide a valuable test of such
models that does not make direct assumptions on viewing angles. 
Observationally, population of the columns around and above 10$^{24}$cm$^{-2}$ 
will be important to characterize the effect of missing the most absorbed
objects in a sample like ours.

\begin{figure}
\includegraphics[width=\columnwidth]{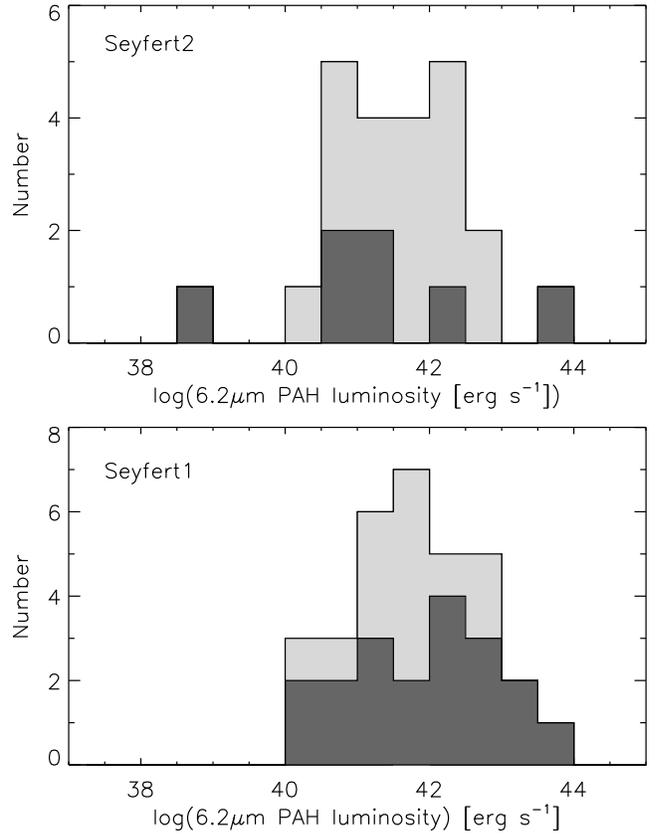}
\caption{Distribution of luminosities of the 6.2$\mu$m PAH emission
feature. Dark shading indicates upper limits.}
\label{fig:pahhisto}
\end{figure}

\begin{figure}
\includegraphics[width=\columnwidth]{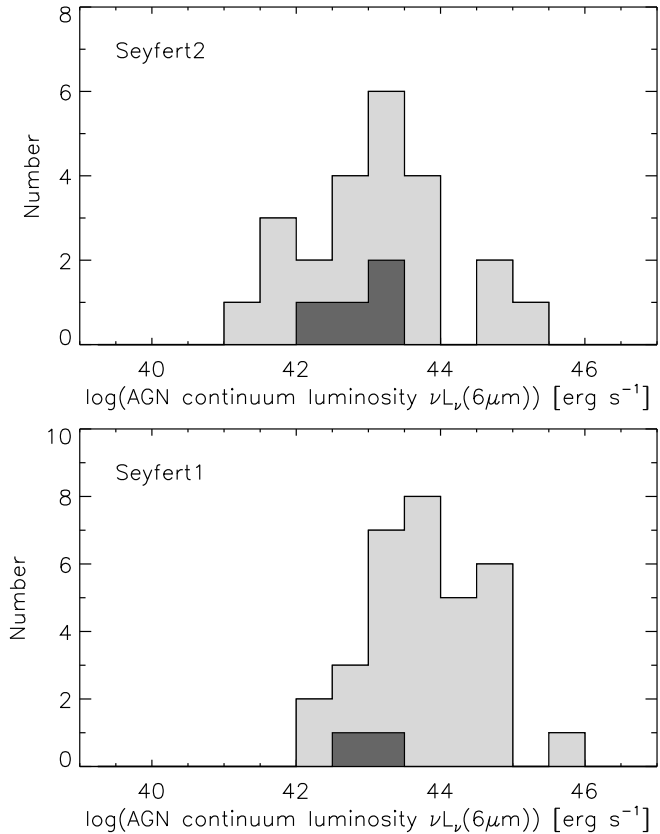}
\caption{Distribution of luminosities of the 6$\mu$m AGN continuum.
Dark shading indicates upper limits.}
\label{fig:conthisto}
\end{figure}

\begin{figure}
\includegraphics[width=\columnwidth]{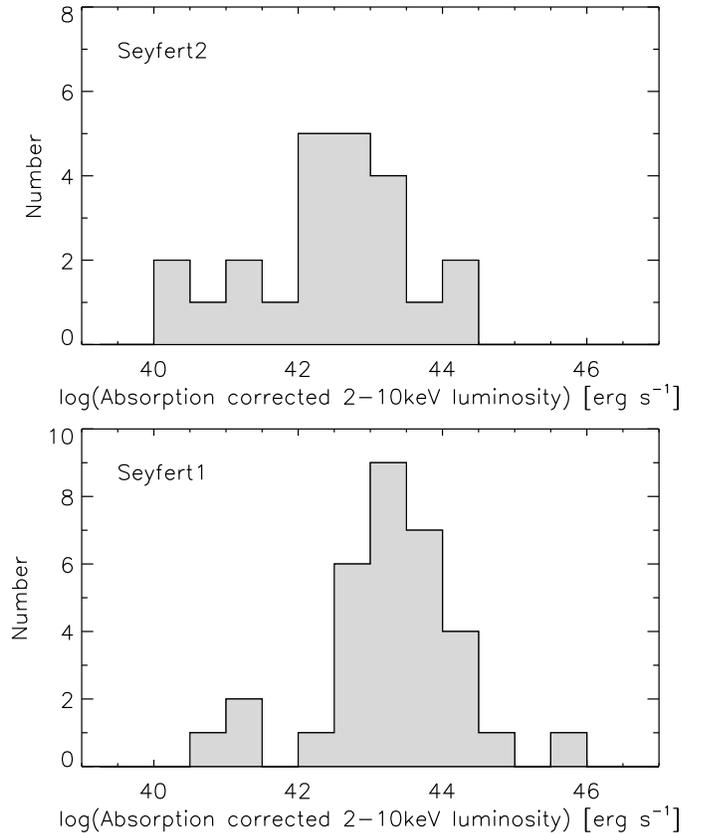}
\caption{Distribution of extinction corrected hard X-ray luminosities.}
\label{fig:xhisto}
\end{figure}

\begin{table*}
\caption{Properties of the Type 1 and Type 2 objects. The
numbers quoted refer to the combination of detections and limits (see
Figs.~\ref{fig:ratiohistogram} to \ref{fig:xhisto} for the fraction of 
limits which is highest among the Sy 1 PAH 
luminosities). Distances refer to the full sample, all others to the part
with extinction corrected hard X-ray data (32 Type 1, 23 Type 2)}
\begin{tabular}{lrrrr}\hline
Property       &\multicolumn{2}{c}{Seyfert 1}&\multicolumn{2}{c}{Seyfert 2}\\
                   &Median   &Disp.  &Median  &Disp.\\ \hline
log(Distance) [Mpc]&2.00     &0.50   &1.68    &0.61 \\  
log(6.2$\mu$m PAH luminosity) [erg s$^{-1}$]&(41.88)&(0.85)&41.50&0.99\\
log(AGN continuum luminosity $\nu$L$_\nu$(6$\mu$m)) [erg s$^{-1}$]
                                            &43.88&0.84&43.20&1.00\\
log(Extinction corrected 2-10keV luminosity) [erg s$^{-1}$]
                                            &43.39&0.99&42.64&1.07\\
log(F(2-10keV)/$\mu$F$_\nu$(6$\mu$m))    &-0.41&0.40&-0.63&0.69\\ 
\hline
\end{tabular}
\label{tab:types}
\end{table*}

One possibility why we fail to observe the expected signature of a torus 
is that much
of the mid-infrared emission is from a more extended region emitting
isotropically, and overwhelming a possible anisotropic emission.
Observations have shown this to be the case for the 10$\mu$m emission
of NGC\,1068 where at least two thirds of the emission comes from an 
extended region overlapping the Narrow Line Region (e.g., Cameron et al.
\cite{cameron93}, Bock et al. \cite{bock98}) rather than a compact
parsec-scale torus. A similar result (at least 27\% extended emission)
has recently been found for NGC\,4151 (Radomski et al. \cite{radomski03}).
High resolution observations and mid-infrared interferometry
with large telescopes should be able to test in the near future whether 
the same applies to other Seyferts. This scenario of (mostly) larger scale and
isotropic AGN continuum at first glance contradicts the results of Clavel
et al. (\cite{clavel00}), however, who find in their large
aperture ISOPHOT-S data a $\sim$8 times lower AGN continuum to
host PAH ratio in Seyfert 2s compared to Seyfert 1s. They ascribed
this to orientation effects in a unified scenario, of a magnitude
significantly larger
than suggested earlier by Heckman (\cite{heckman95}) and
Maiolino et al. (\cite{maiolino95}). Much of their result
could be an AGN luminosity effect rather than an orientation effect, though. 
Some 
of the relevant diagnostics are shown in Figs.~\ref{fig:pahhisto} to 
\ref{fig:xhisto} and Table~\ref{tab:types}
for our sample which has a significant overlap with the Clavel et al. 
(\cite{clavel00}) sample. The distributions of
PAH luminosities for the two types are consistent (Fig.~\ref{fig:pahhisto}),
but the caveat that only upper limits to the 6.2$\mu$m PAH are 
available for many of the strong continuum Sy1s has to be noted. 
This result means there are no obvious differences in the host properties
as probed by the PAH emission, although some differences (as suggested
e.g. by Maiolino et al. \cite{maiolino95}) might be hidden
by the many PAH upper limits, or by sample and physical aperture size 
(distance) mismatch between the two types.
We note that upper limits also apply to the 6.2 and 
7.7$\mu$m PAH features in many objects presented by Clavel et al. 
(\cite{clavel00}), rather than detections reported through their method of 
including only positive flux above the noisy continuum into their 
PAH fluxes. For our sample, both the mid-infrared continuum luminosities
(Fig.~\ref{fig:conthisto}) and the absorption corrected X-ray luminosities
(Fig.~\ref{fig:xhisto}) are larger in Seyfert 1s than in Seyfert 2s. 
Reflecting the luminosity differences, a difference is also seen
in the distance distributions (Fig.~\ref{fig:disthisto}). All that
argues that a difference in PAH equivalent width between Seyfert 1s and 2s, 
found by Clavel et al.
(\cite{clavel00}), and also seen in our data (compare Fig.~\ref{fig:pahhisto}
and Fig.~\ref{fig:conthisto}), can be largely due to the
Seyfert 1 part of the samples representing both higher AGN luminosities
and higher AGN/host luminosity ratios. This is a natural consequence of many
AGN samples, including ones driving the selection of ISO observations, being
biased in favour of more distant Type 1 objects (see discussion in 
Maiolino \& Rieke \cite{maiolino95b}).
Explaining our results by a dominant
component of isotropic mid-infrared emission thus remains viable. We emphasise
that a minor strongly anisotropic mid-infrared component may be present 
without affecting our analysis. 

Another factor masking the signature of unification in the mid-IR to
hard X-ray ratios could be uncertainties in the absorption correction
of the X-ray fluxes. While we have argued above that they may contribute
to the increased scatter observed for the Type 2 objects, we see no
evidence for a systematic offset, which would also have profound implications
on estimates of the energy budget of obscured X-ray sources. 
Similar arguments apply to scatter induced by intrinsic variations of the
AGN spectral energy distributions between the hard X-rays and the wider 
spectral range that is heating dust.
Finally, the strategy of our comparison relies on the assumption
of isotropic X-ray emission (once considering the absorption corrections).
If X-rays were emitted anisotropically and if there were alignment between
the structures causing this X-ray anisotropy and a torus absorbing/emitting
anisotropically in the mid-IR, the expected difference in the
ratio of hard X and mid-IR emission could cancel largely.

Our results suggest that local, mostly moderate luminosity, Seyferts 
do not show the behaviour expected if their mid-infrared emission were 
dominated by a compact, 
anisotropically emitting torus. This adds to evidence that the most
simple version of a unified scheme, in which all AGN at all luminosities
are governed by such a structure, is not applicable. Recent deep X-ray 
and mid-IR surveys indicate significant differences in the evolution
of high luminosity Type 1 and lower luminosity Type 2 populations that 
cannot be reconciled with a single type of obscuring structure in AGN
of all luminosities and redshifts (e.g. Franceschini et al. 
\cite{franceschini02}).

\section{Conclusions}

We have used spectral decomposition of a large sample of ISO spectra of AGNs
to isolate the AGN 6$\mu$m continua. We compare these mid-infrared continua
to intrinsic hard X-ray fluxes from the literature, assumed to be a fair
isotropic measure of AGN luminosity. Due to this normalization, our comparison
can test AGN properties and unification aspects without the level of 
sensitivity to selection biases (e.g. luminosities, ratio of Type 1 and Type 2 
objects in the sample) that is found in comparisons of absolute quantitities 
or of normalizations to non-AGN quantitities. The main results are:

(1) Mid-infrared and intrinsic X-ray fluxes correlate, but with significant
scatter. Ratios vary by more than an order of magnitude, the dispersion in 
the log is 0.4 for Seyfert 1 and 0.69 for Seyfert 2 galaxies. Main 
contributors to this 
spread likely include variations in the AGN spectral energy distribution and 
geometry of obscuring dust, as well as the effects of AGN variability.

(2) There is no significant difference between Type 2 and Type 1 in the 
average ratio of X-ray and mid-infrared continuum, in contrast to 
expectations from unified scenarios invoking an optically and geometrically 
thick torus emitting anistropically in the mid-IR. Most likely, this is due
to a large contribution from extended dust components emitting more 
isotropically. This will dilute anisotropic emission that, however, may still 
be present at lower levels.

\begin{acknowledgements} We thank Sue Madden for information on the mid-IR
    properties of elliptical galaxies, the referee for helpful comments and 
    Eckhard Sturm for discussions.
    We acknowledge support for the ISO spectrometer data center at MPE by DLR 
   (50 QI 0202).
\end{acknowledgements}

\appendix
\section{Notes on selected objects}

\noindent {\bf NGC\,1808} is believed to host both a strong spatially
extended starburst (e.g. Krabbe et al. \cite{krabbe94}) and possibly a 
weak and likely fading AGN (e.g. Bassani et al. \cite{bassani99}),
although the X-ray luminosity is low enough to overlap the regime of 
(also variable) non-AGN 
Ultraluminous X-ray Sources (ULX). Such objects are  found at non-nuclear 
positions in other starbursting objects (Fabbiano et al. 
\cite{fabbiano03}). Because of the combination of strong star formation 
and weak AGN, the limit on AGN continuum
is far off the correlation between mid-IR and X-rays.

\noindent{\bf NGC\,1667} is likely another example of a fading `fossil'
AGN (Bassani et al. \cite{bassani99}), introducing additional uncertainty 
on the AGN contribution at different wavelengths. For example, if the
AGN is turning off the apparently high $N_H$ may be an artifact of the 
reflected component remaining visible longer. 

\noindent {\bf NGC\,4945}. The 6$\mu$m continuum is high compared to
the average starburst, but not at the level of a significant detection
of the AGN continuum given the scatter in starburst properties. The 
mid-IR weakness and corresponding location in our diagrams is 
certainly in part due to
the foreground extinction which is high even in the mid-infrared 
in this system (Spoon et al. \cite{spoon00}). A high resolution
mid-IR image, if possible outside the silicate band, is needed to break 
the ambiguity of both this result
and the low resolution mid-IR imaging of Krabbe et al. (\cite{krabbe01}),
and reach a mid-IR detection of the AGN.

\noindent {\bf NGC\,6240} is one of the sources deviating most from the 
correlation, in the sense of high X-ray and low mid-IR. This could include a 
contribution of adopting a high estimate for the intrinsic X-ray emission 
(cf. also the lower estimate of Ikebe et al. \cite{ikebe00}), and the 
likely noticeable 
obscuration of the 6$\mu$m continuum in this object. See Lutz et al. 
(\cite{lutz03}) for a full discussion of the mid-infrared properties.

\noindent {\bf Mrk\,463} has a very strong mid-IR AGN continuum but the 
relatively lowest absorption corrected hard X-ray emission compared to the 
average relation. The X-ray column may be underestimated.


\begin{thebibliography}{}

\bibitem[2001]{alonso01} Alonso-Herrero, A., Quillen, A.C., Simpson, C.,
Efstathiou, A., Ward, M.J. 2001, \aj, 121, 1369

\bibitem[1993]{antonucci93} Antonucci, R. 1993, \araa, 31, 473

\bibitem[1999]{bassani99} Bassani, L., Dadina, M., Maiolino, R., et al.
1999, \apjs, 121, 473

\bibitem[2002]{blustin02} Blustin, A.~J., 
Branduardi-Raymont, G., Behar, E., et al.\ 2002, \aap, 392, 453

\bibitem[1998]{bock98} Bock, J.J., Marsh, K.A., Ressler, M.E., Werner, M.W.
1998, \apj, 504, L5 

\bibitem[2002]{boller02} Boller, T., Gallo, L.C., Lutz, D., Sturm, E.
2002, \mnras, 336, 1143

\bibitem[2003]{braito03} Braito, V.~et al.\ 2003, \aap, 398, 107

\bibitem[1997]{brandt97} Brandt, W.~N., Fabian, 
A.~C., Takahashi, K., et al.\ 1997, \mnras, 290, 617

\bibitem[1993]{cameron93} Cameron, M., Storey, J.W.V., Rotaciuc, V., et al.
1993, \apj, 419, 136

\bibitem[1999]{cappi99} Cappi, M.~et al.\ 1999, \aap, 344, 857

\bibitem[2000]{clavel00} Clavel, J., Schulz, B., Altieri, B., et al. 2000,
\aap, 357, 839

\bibitem[2001]{dellaceca01} Della Ceca, R., 
Pellegrini, S., Bassani, L., et al. \ 2001, \aap, 375, 781

\bibitem[2002]{dellaceca02} Della Ceca, R., Ballo, L., Tavecchio, F.,
et al. 2002, \apj, 581, L9

%\bibitem[1995]{efstathiou95} Efstathiou, A., Hough, J.H., Young, S. 1995,
%\mnras, 277, 1134
\bibitem[1995]{efstathiou95} Efstathiou, A., Rowan-Robinson, M. 1995,
\mnras, 273, 649

\bibitem[2003]{fabbiano03} Fabbiano, G., Zezas, A., King, A.R., et al.
2003, \apj, 584, L5

\bibitem[2002]{franceschini02} Franceschini, A., Braito, V., Fadda, D.
2002, \aap, 335, L51

\bibitem[2001]{fukazawa01} Fukazawa, Y., Iyomoto, 
N., Kubota, A., Matsumoto, Y., Makishima, K.\ 2001, \aap, 374, 73

\bibitem[1995]{genzel95} Genzel, R., Weitzel, L., Tacconi-Garman, L.E.,
Blietz, M., Cameron, M., Krabbe, A., Lutz, D. 1995, \apj, 444, 129

%\bibitem[1998]{genzel98} Genzel, R., Lutz, D., Sturm, E., et al. 1998, 
%\apj, 498, 579

\bibitem[2000]{george00} George, I.~M., Turner, 
T.~J., et al.\ 2000, \apj, 531, 52

\bibitem[2001a]{gondoin01a} Gondoin, P., Lumb, D., 
Siddiqui, H., Guainazzi, M., Schartel, N.\ 2001a, \aap, 373, 805

\bibitem[2001b]{gondoin01b} Gondoin, P., Barr, P., 
Lumb, D., et al.\ 2001b, \aap, 378, 806

\bibitem[2002]{gondoin02} Gondoin, P., Orr, A., Lumb, D., Santos-Lleo, M. 
2002, \aap, 388, 74

\bibitem[2003a]{gondoin03a} 
Gondoin, P., Orr, A., Lumb, D., Siddiqui, H.\ 2003a, \aap, 397, 883

\bibitem[2003b]{gondoin03b} Gondoin, P., 
Orr, A., \& Lumb, D.\ 2003b, \aap, 398, 967

\bibitem[1997]{granato97} Granato, G.L., Danese, L., Franceschini, A. 1997,
\apj, 486, 147

\bibitem[2000]{guainazzi00} Guainazzi, M., Matt, 
G., Brandt, W.~N., et al.\ 2000, \aap, 356, 463

\bibitem[2001]{guainazzi01} 
Guainazzi, M., Marshall, W., Parmar, A.~N.\ 2001, \mnras, 323, 75

\bibitem[1995]{heckman95} Heckman, T.M. 1995, \apj, 446, 101

%\bibitem[1997]{ho97} Ho, L.C., Filippenko, A.V. 1997, \apjs, 112, 315

%\bibitem[1999]{ho99} Ho, L.C. 1999, \apj, 516, 672

\bibitem[1992]{huchra92} Huchra J., Burg R., 1992, \apj, 393, 90

\bibitem[2000]{ikebe00} Ikebe, Y., Leighly, K., Tanaka, Y., et al. 2000,
\mnras, 316, 433

\bibitem[2003]{immler03} Immler, S., Brandt, 
W.~N., Vignali, C., et al.\ 2003, \aj, 126, 153

\bibitem[1996]{iyomoto96} Iyomoto, N., Makishima, 
K., Fukazawa, Y., et al. 1996, \pasj, 48, 231

\bibitem[1994]{krabbe94} Krabbe, A., Sternberg, A., Genzel, R. 1994,
\apj, 425, 72

\bibitem[2001]{krabbe01} Krabbe, A., B\"oker, T., Maiolino, R. 2001, \apj,
557, 626

\bibitem[2001]{landi01} Landi, R.~et al.\ 2001, \aap, 379, 46

\bibitem[2000]{laurent00} Laurent, O., Mirabel, I.F., Charmandaris, V., et al.
2000, \aap, 359, 887 

\bibitem[1997]{lawson97} Lawson, A.~J., Turner, M.~J.~L. 1997, \mnras, 288, 920

\bibitem[2003]{lu03} Lu, N., Helou, G., Werner, M.W., et al. 2003, \apj, 
588, 199

\bibitem[2003]{lutz03} Lutz, D., Sturm, E., Genzel, R., et al. 2003,
\aap, 409, 867 

\bibitem[1995]{maiolino95} Maiolino, R., Ruiz, M., Rieke, G.H., Keller, L.D.
1995, \apj, 446, 561

\bibitem[1995]{maiolino95b} Maiolino, R., Rieke, G.H. 1995, ApJ, 454, 95

\bibitem[1998]{maiolino98} Maiolino, R., Salvati, M., Bassani, L., 
et al. 1998, \aap, 338, 781

\bibitem[1997]{malizia97} Malizia, A., Bassani, 
L., Stephen, J.~B., Malaguti, G., \& Palumbo, G.~G.~C.\ 1997, \apjs, 113, 
311

\bibitem[1999]{malizia99} Malizia, A., Bassani, L., Zhang, S.N., et al. 
1999, \apj, 519, 637

\bibitem[1997]{matt97} Matt, G.~et al.\ 1997, \aap, 325, L13

\bibitem[1999]{matt99} Matt, G.~et al.\ 1999, \aap, 341, L39

\bibitem[2000]{matt00} Matt, G., Perola, G.~C., 
Fiore, F., Guainazzi, M., Nicastro, F., Piro, L.\ 2000, \aap, 363, 863

\bibitem[1994]{mazzarella94} Mazzarella, J., Voit, G.M., Soifer, B.T., 
Matthews, K., Graham, J.R., Armus, L., Shupe, D.L. 1994, \aj, 107, 1274
 
\bibitem[1998]{mirabel98} Mirabel, I.F., Vigroux, L., Charmandaris, V.,
et al. 1998, \aap, 333, L1

\bibitem[1994]{nandra94} Nandra, K.~ Pounds, K.~A.\ 1994, \mnras, 268, 405

\bibitem[2002]{nenkova02} Nenkova, M., Ivezi\'c, Z., Elitzur, M. 2002,
\apj, 570, L9 

\bibitem[2000]{perola00} Perola, G.~C.~et al.\ 2000, \aap, 358, 117

\bibitem[2002]{perola02} Perola, G.~C., Matt, G., 
Cappi, M., et al.\ 2002, \aap, 389, 802

\bibitem[1982]{piccinotti82} Piccinotti, G. , Mushotzky, R.F., Boldt, E.A., 
et al. 1982, ApJ, 253, 485
 
\bibitem[1992]{pier92} Pier, E.A., Krolik, J.H. 1992, \apj, 401, 99

\bibitem[1996]{polletta96} Polletta, M., Bassani, L., Malaguti, G., 
Palumbo, G.~G.~C., Caroli, E.\ 1996, \apjs, 106, 399 

\bibitem[2001]{obrien01} O'Brien, P.T., et al. 2001, \mnras, 327, L37

\bibitem[1997]{ogasaka97} Ogasaka, Y., Inoue, H., 
Brandt, W.~N., et al.\ 1997, \pasj, 49, 179

\bibitem[2003]{radomski03} Radomski, J.T., Pi\~na, R.K., Packham, C., et al.
2003, ApJ, 587, 117

\bibitem[1992]{rao92} Rao, A.~R., 
Singh, K.~P., Vahia, M.~N.\ 1992, \mnras, 255, 197

\bibitem[2000]{reeves00} Reeves, J.~N., Turner, M.~J.~L.\ 2000, \mnras, 
316, 234 

\bibitem[1997]{reynolds97} Reynolds, C.~S.\ 1997, \mnras, 286, 513

\bibitem[2000]{risaliti00} 
Risaliti, G., Maiolino, R., \& Bassani, L.\ 2000, \aap, 356, 33 

\bibitem[2001]{severgnini01} Severgnini, P., 
Risaliti, G., Marconi, A., Maiolino, R., Salvati, M.\ 2001, \aap, 368, 44

\bibitem[1995]{spinoglio95} Spinoglio, L., Malkan, M.A., Rush, B.,
Carrasco, L., Recillas-Cruz, E. 1995, \apj, 453, 616

\bibitem[2000]{spoon00} Spoon, H.W.W., Koornneef, J., Moorwood, A.F.M.,
Lutz, D., Tielens, A.G.G.M. 2000, \aap, 357, 898

\bibitem[2002]{spoon02} Spoon, H.W.W., Keane, J.V., Tielens, A.G.G.M., et al.
2002, \aap, 385, 1022

\bibitem[2003]{spoon03} Spoon, H.W.W., et al. 2003, \aap, in press 
(astro-ph/0310721)

\bibitem[2000]{sturm00} Sturm, E., Lutz, D., Tran, D., et al. 2000,
\aap, 358, 481 

\bibitem[1999]{thuan99} Thuan, T.X., Sauvage, M., Madden, S. 1999, \apj,
516, 783

\bibitem[2001]{tran01} Tran, Q.D., Lutz, D., Genzel, R., et al. 2001,
\apj, 552, 527

\bibitem[2001]{turner01} Turner, T.~J., Romano, 
P., George, I.~M., et al.\ 2001, \apj, 561, 131

\bibitem[1999]{vignati99} Vignati, P.~et al. 1999, \aap, 349, L57

\bibitem[2003]{xilouris03} Xilouris, E.M., Madden, S.C., Galliano, F., 
Vigroux, L., Sauvage, M. 2003, \aap, in press (astro-ph/0312029)

\bibitem[2001]{yang01} Yang, Y., Wilson, A.~S., Ferruit, P.\ 2001, 
\apj, 563, 124

\end{thebibliography}
\end{document}